\documentclass[useAMS,usenatbib]{mn2e}
\usepackage{graphicx}	
\usepackage{lineno}	
	
\title[Differential coronal rotation using radio images at 17 GHz]{Differential coronal rotation using radio images at 17 GHz}

\author[Satish Chandra, Hari Om Vats and K. N. Iyer]{Satish Chandra$^{1}$\thanks{
satish0402@gmail.com (SC),} Hari Om Vats$^{2}$\thanks{vats@prl.res.in (HOV),} and K. N.
Iyer$^{3}$\thanks{iyerkn@yahoo.com (KNI)}\\
$^{1}$Department of Physics, PPN College, Kanpur - 208 001, INDIA.\\
$^{2}$Physical Research Laboratory, Ahmedabad - 380 009, INDIA.\\
$^{3}$Department of Physics, Saurastra University, Rajkot - 360 005, INDIA.}
\begin{document}


\pagerange{\pageref{firstpage}--\pageref{lastpage}} \pubyear{8888}

\maketitle


\label{firstpage}

\begin{abstract}
In the present work, we perform time-series analysis on the latitude bins of the solar full disk (SFD) images of Nobeyama Radioheliograph ({\it NoRH\/}) at 17 GHz. The flux modulation method traces the passage of radio features over the solar disc and the autocorrelation analysis of the time-series data of SFD images (one per day) for the period 1999--2001 gives the rotation period as a function of latitude extending from 60\degr S to 60\degr N. The results show that the solar corona rotates less differentially than the photosphere and chromosphere, {\it i.e.}, it has smaller gradient in the rotation rate. 
\end{abstract}

\begin{keywords}
Sun: corona -- Sun: radio radiation -- Sun: rotation
\end{keywords}

\section{Introduction}

In the recent studies of the solar rotation, the tracer methods were extensively used for the determination. The features like sunspots, plages, filaments, faculae, bright points, super granules, coronal holes, giant cells, etc. were found to rotate with the Sun \citep{b8,b9,b15}. By tracing the passage of any of these features over the solar disc, one can estimate the rotation as well as its differential nature with respect to height and latitude. In general, the equatorial region rotates faster than the polar-region.

The solar coronal rotation is a comparatively less understood phenomenon because the coronal features are less prominent in both the time duration and spatial extent. The observations are affected by the fact that the solar corona is optically thin across a wide range of observed radio frequencies. Recent studies of the solar radio emission provided a lot of information concerning the solar corona \citep{b16, b12}. The radio observations have the advantage of permitting one to attain resolution for the heights in the corona as the propagation of radio waves distinctly depends on the plasma density in the corona. By using radio flux data at different frequencies \citep{b17}, it has been shown earlier that the coronal rotation depends on the heights in the corona.

Several groups have reported from time to time that the solar corona rotates differentially, with respect to latitude, in the same way as the photosphere and the chromosphere. The analysis of X-ray bright points (XBPs) in {\it SOHO}/EIT images \citep{b4,b5,b6,b10}, and recent analysis of XBPs in {\it Hinode}/XRT and {\it Yohkoh}/SXT  images \citep{b11} also asserts the coronal differential rotation. However, in contrast to these references, \citet{b18} and \citet{b19} studied the rotation rate of corona by using {\it Yohkoh}/SXT data and showed that the soft X-ray corona does not exhibit any significant differential rotation, {\it i.e.} the corona rotates more rigidly than the chromosphere and the photosphere.

Recently, by making use of XBPs observed with the {\it Yohkoh}/SXT, \citet{b21} found that the rate of differential rotation changes with a parameter $\triangle t$ which is linked with the lifetime of XBPs. The long-lived XBPs follows the rotation rate of the photospheric magnetic field, whereas, the short-lived XBPs approaches to that evaluated from the photospheric Doppler measurements.

\citet{b20} compared differential rotation of subphotospheric layers derived from GONG++ dopplergrams with the small bright coronal structure (SBCS) observed through {\it SOHO}/EIT for the period correspond to the decling phase of solar cycle 23. They reported that the SBCS rotate faster than the upper subphotospheric layer (3Mm) by about 0.5 deg/day at the equator. The latitude gradient of the rotation rate of the SBCS and the subphotospheric layers were found to be very similar.

The comparison of the data of the solar magnetic elements reveal that the differential rotation of the compact magnetic elements with negative and positive polarities have similar behavior for the solar cycles 20 and 21 \citep{b22}. It is also established that some variations in the rotation rate of compact magnetic elements were present at the time of polarity reversal of the Sun.

The differential rotation of the solar corona is, therefore, still an open subject and further studies are required to establish the coronal differential rotation. Here we report the study of the coronal rotation using Nobeyama Radioheliograph ({\it NoRH\/}).

\begin{figure*}
  \centering{
  \includegraphics[width=\textwidth]{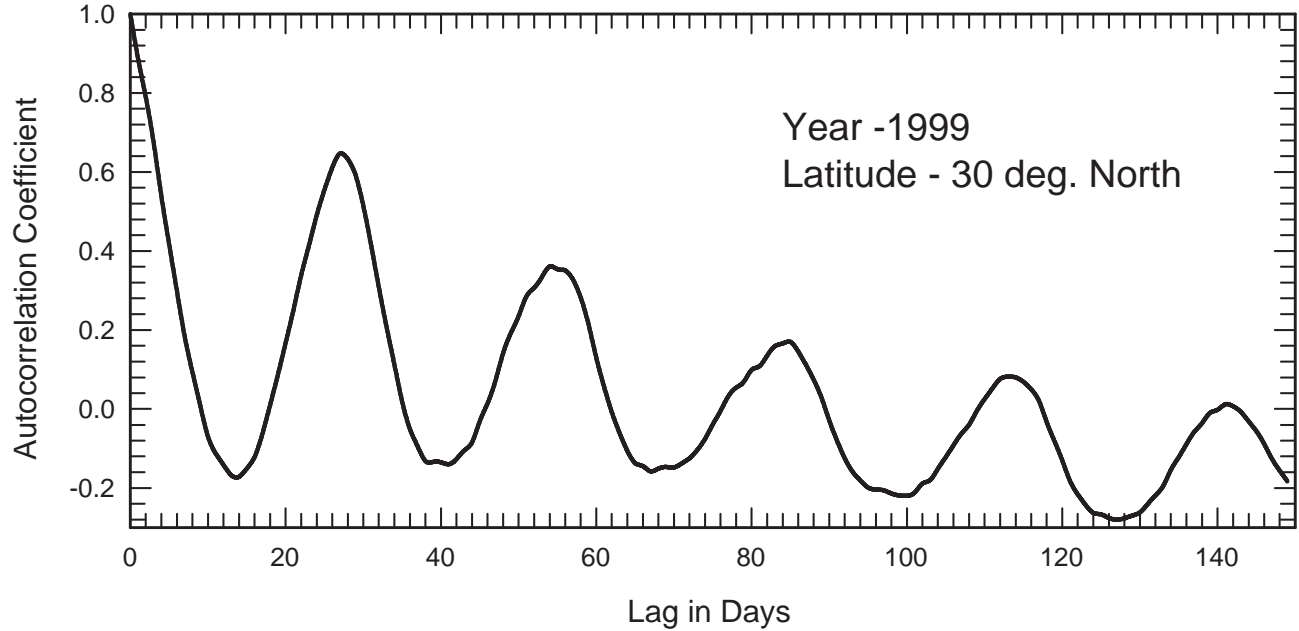}}
 \caption{Typical autocorrelogram of the year 1999, used in the present work for the estimation of synodic rotation period.}
 \end{figure*}

\section{Observations and Methodology}

The Nobeyama Radioheliographs at 17 GHz consist of an array of eighty four antennas aligned in the T-shaped configuration. This is dedicated only for the solar observations. High speed processing of signals from all antennas generates full disc radio images of the Sun, at the maximum rate of 20 images per second. The Sun is observed for 8 hours per day. The radioheliograph provides unprecedented uniform sampling of SFD images with high spatial and temporal resolutions of 10 arcsec and 50 msec, respectively \citep{b14}. The observations are possible throughout the year without any interruption by weather conditions. The radio images seem to discriminate low-contrast or faint structures on the solar disk {\it e.g.} dark filaments, bright points, coronal holes, and other large-scale structures. 

Here, we analyze the solar full disk images obtained from Nobeyama Radioheliograph ({\it NoRH\/}) at 17 GHz. Each image is of 512 x 512 pixel in size. The principal technique of our investigation is to calculate the autocorrelation coefficient of the intensity variation for a set of latitude bins selected on the {\it NoRH\/} images. For this we utilize one {\it NoRH\/} full disk image per day on which latitude bins, or strips, are chosen. The dimensions of these rectangular bins are just two pixels in width and the length is such as to incorporate the entire pixels on the solar disc image at each latitude. The strips are selected at every 10\degr latitude

For each latitude, a time series of the radio intensity is generated from the average of the  selected pixels values in the strip. To determine rotation period, we obtained autocorrelation coefficient using the standard subroutines of IDL software. The period estimation is a straightforward identification of any dominating periodically varying signal present in the coronal radio emission.

\begin{table}
 \centering
 \begin{minipage}{75mm}
  \caption{The coefficient $A$ \& $B$ obtained from Nobeyama Radioheliograph (NoRH) at 17 GHz for the years 1999-2000. The last column gives annual sunspot numbers.}
  \begin{tabular}{@{}llrrl@{}}
  \hline
   Years 			& \multicolumn{2}{c}{Coefficient} 						&	Annual\\
     						& $A \pm E_A $    		&  $B \pm E_B  $ 				&	Sunspot\\       
  							&											&												& Numbers\\
  \hline
    1999				&	14.826 $\pm$ 0.200	&	$-$1.528 $\pm$ 0.474	&	093.3 \\
    2000				&	15.083 $\pm$ 0.239	&	$-$2.892 $\pm$ 0.567	&	119.6 \\
    2001				&	14.543 $\pm$ 0.097	&	$-$1.968 $\pm$ 0.229	&	111.0 \\
   \textbf{Mean}& 14.817 $\pm$ 0.060  & $-$2.129 $\pm$ 0.143	& 107.9 \\
  \hline
\end{tabular}
\end{minipage}
\end{table}

\section{Data Analysis}

For the time-series of each latitude, autocorrelation upto a lag of 145 days are calculated and plotted in the form of autocorrelogram. The typical autocorrelogram for a data set for the year 1999 at 30\degr north latitude is shown in Figure~1. A systematic periodic modulation can be seen from the autocorrelogram as a modulation due to the solar rotation. The autocorrelation coefficient at some of the latitudes are quite high and the curve is quite smooth showing many damped cyclic oscillations (Figure~1). The amplitude of oscillatory features reduce perhaps due to the temporal variability in the solar rotation. This is because of the long term persistence of the solar features on the solar disc at that latitude. The short term features at higher latitudes affect adversely the smoothness, numbers of cyclic variation as well as the magnitude of autocorrelation coefficient. Nevertheless, the first peak of each autocorrelogram is quite clear and could be used for the estimation of the synodic rotation period upto $\pm 60\degr$ latitude. The average radio flux at higher latitudes ($>60\degr$), in both the hemispheres, does not show periodic nature and hence is not used for obtaining rotational periods. Thus from {\it NoRH\/} images we are able to get rotational periods from 60\degr S to 60\degr N only. The radio emission at higher latitudes is either uniform or totally random as a function of time. Employing this method, we determined first the synodic rotation period and then from the synodic the sidereal rotation period, at every 10\degr latitude in the range of $\pm 60\degr$ each year during 1999-2001.

To compare the sidereal rotation rate $\Omega(\psi)$ of this work with other observations and their models of the solar rotation, we fit the data using the standard polynomial expansion in its most commonly used form, as given below

\begin{equation}
\Omega(\psi)=A + B \sin^2\psi
\end{equation}

where $\Omega(\psi)$ is the sidereal rotation rate at heliographic latitude $\psi$. The parameter $A$ represents the equatorial rotation rate and $B$ represents differential rate largely at low latitudes. Since we have not used the data of the higher latitudes $>60\degr$ on both the sides of the equator, hence, it is sufficient to use only the first two terms of the standard polynomial expansion \citep{b3}.

The rotation rate is related to the rotation period as given below,

\begin{equation}
T(\psi)=\frac {360\degr}{\Omega(\psi)}
\end{equation}

where $T(\psi)$ is in day and $\Omega(\psi)$ in degree/day.

\begin{figure}
  \resizebox{\hsize}{!}{
  \includegraphics{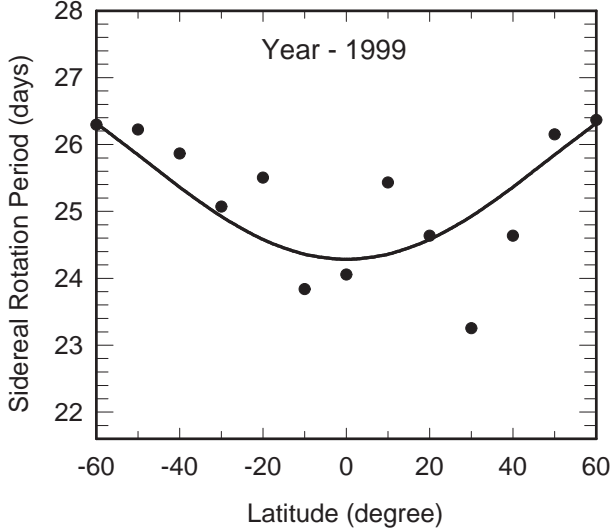}}
 \caption{The coronal sidereal rotation period as a function of the latitude for the year 1999}
\end{figure}

\begin{figure}
\resizebox{\hsize}{!}{
  \includegraphics{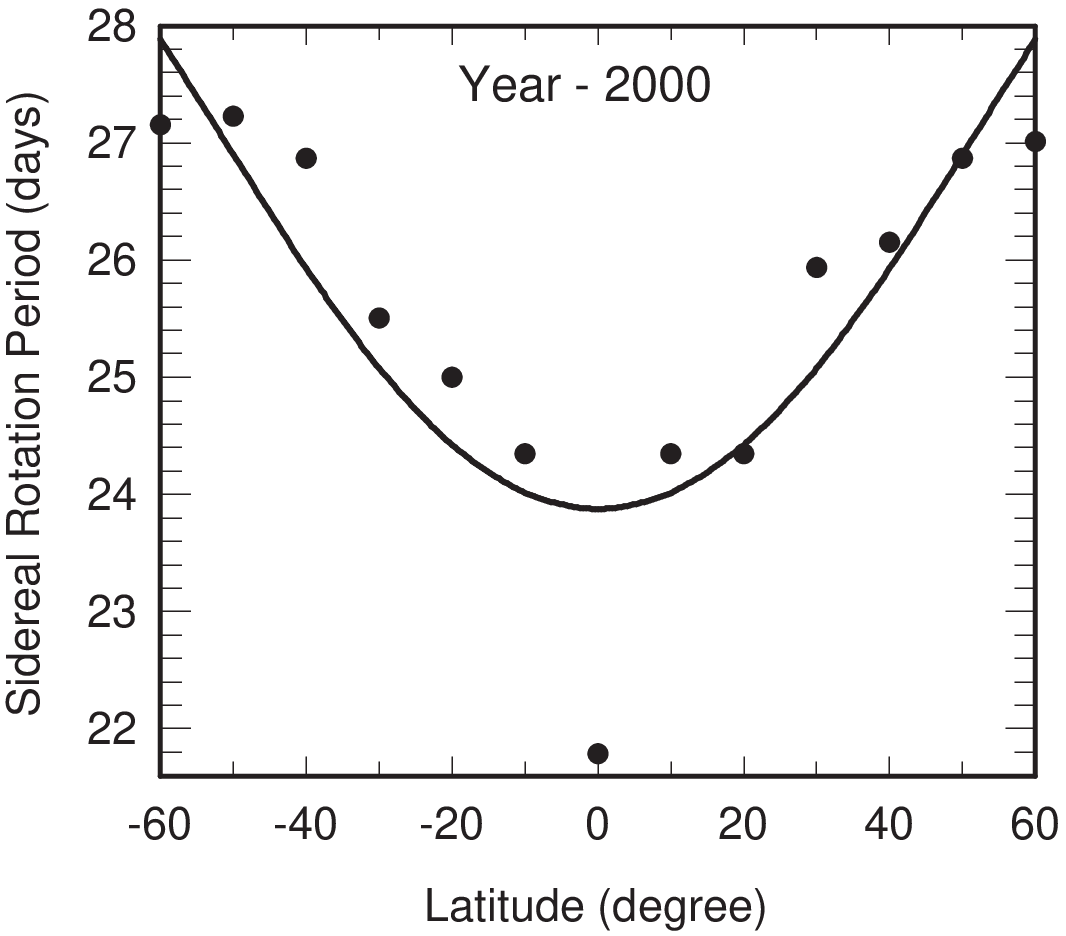}}
    \caption{As in Figure~2 for the year 2000}
\end{figure}

\begin{figure}
 \resizebox{\hsize}{!}{
  \includegraphics{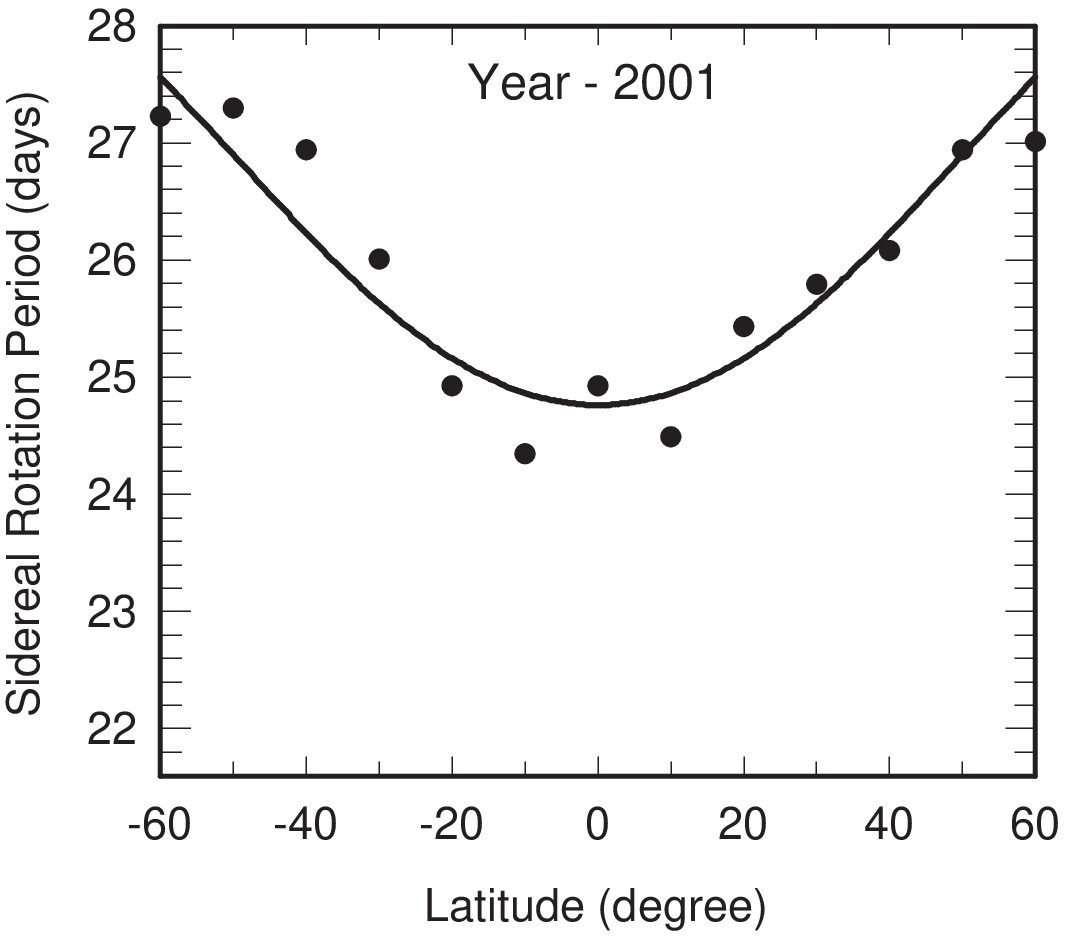}}
 \caption{As in Figure~2 for the year 2001}
\end{figure}

\begin{table*}
 \centering
 \begin{minipage}{160mm}
 \caption{The coefficients $A$ \& $B$ obtained from the present work and other recent published results.}
 \begin{tabular}{@{}lllrrll@{}}
 \hline
  Data Source & Tracers 						& \multicolumn{2}{c}{Coefficients} 		& References 	&Period \\
              &	 	      						& $A \pm E_A$    	&$B \pm E_B$ 				&             &\\
 \hline
  NoRH			 	&flux modulation 		  &14.817$\pm$0.060	&$-$2.129$\pm$0.143	&present work	&1999-2001 \\
  Hinode/XRT	&X-ray bright points	&14.192$\pm$0.170	&$-$4.211$\pm$0.775	&\citet{b11}	&Jan, Mar \& Apr 2007 \\
  Yohkoh/SXT	&Coronal bright points&17.597$\pm$0.398	&$-$4.542$\pm$1.136	&\citet{b11}	&1992-2001 \\
  SOHO/EIT		&Coronal bright points&14.677$\pm$0.033	&$-$3.100$\pm$0.140 &\citet{b6}		&Jun 1998-May 1999 \\
	Greenwich		&Sunspots							&14.551$\pm$0.006	&$-$2.870$\pm$0.060	&\citet{b2}		&1874-1976\\
 \hline
\end{tabular}
\end{minipage}
\end{table*}

\section{Comparison}

The estimated sidereal rotation period has been plotted as a function of the latitude for the years 1999-2001, separately. Figures~2 --~4 represent the differential rotation profiles of each year. In spite of scatter in the data points, the solar differential rotation is clearly evident. We obtained the coefficient $A$ and $B$ of the solar coronal rotation for each year separately. These are tabulated in Table ~1. The mean of these has been compared with the results obtained through the analysis of {\it Hinode}/XRT XBPs, {\it Yohkoh}/SXT XBPs \citep{b11} and {\it SOHO}/EIT CBPs data \citep{b6}. 

The sidereal equatorial rate (coefficient $A$, in Table~2) in the case of {\it NoRH} is found to be 4.4 \% higher than the {\it Hinode}/XRT XBPs (14.2 deg/day) and 15.8 \% lower than the {\it Yohkoh}/SXT XBPs data (17.6 deg/day), on average. Coefficient $A$ of the present work is found to be comparable with the sunspot data (14.6 deg/day) and {\it SOHO}/EIT CBPs data (14.7 deg/day). The differential rate represented by coefficient $B$ is found to be significantly lower then the values obtained by \citet{b11} using {\it Hinode}/XRT XBPs data (49.4 \%) \& {\it Yohkoh}/SXT XBPs data (53.1 \%), \citet{b6} using {\it SOHO}/EIT CBPs data (31.3 \%) and \citet{b2} using photospheric sunspots data (25.8 \%). It means the latitude gradient of the coronal rotation rate is much lower at 17 GHz radio frequency in comparison to that obtained from the observations of X-Ray and EUV frequencies.

From the analysis of radio flux at 2.8 GHz, \citet{b12} could not find any systematic relationship between coronal rotation period and phases of solar cycle. Our results show comparatively faster equatorial rotation in the year 2000, than those of 1999 and 2001. Thus, it may have some relation with the maxima of the solar activity in the 23d cycle. The sunspot numbers in 1999 and 2001 are relatively lower than 2000. The estimated coefficients $A$ and $B$ from the Nobeyama radioheliogram show a temporal variability. The temporal variations of $A$ and $B$ and sunspot numbers are shown in Figure~5.

\begin{figure}
 \resizebox{\hsize}{!}{
  \includegraphics{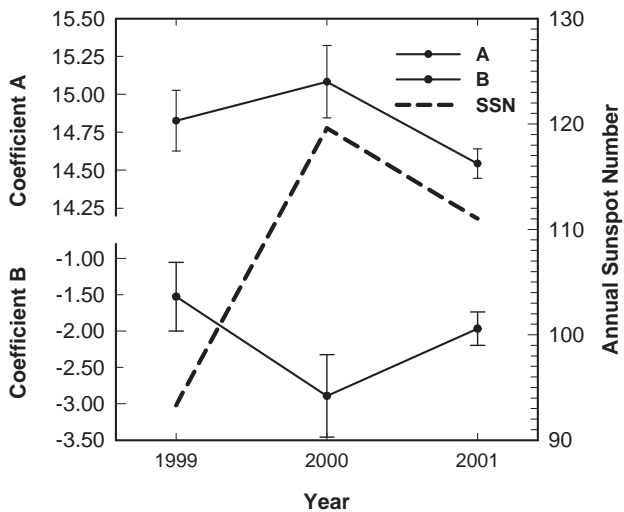}}
 \caption{The coefficients A, B and annualy averaged sunspot numbers for 1999-2000.}
\end{figure}

\section{Conclusion}

Our results for coefficient $A$ are more in agreement with the results derived from the sunspot region at the photospheric \citep{b7,b2,b13} and at the chromospheric level \citep{b6,b10}. Since {\it Yohkoh}/SXT observed the soft X-ray emission of the corona and was not a coronograph, the portion of corona that was imaged through this SXT range from the photosphere to a height of approximately 0.5 $R_{\sun}$ (David McKanzie, Private communication 2009). Whereas, the radio emissions at 17 GHz originate approximately at the altitude of $\sim 1.2 \times 10^4$ km above the photosphere. The height is adopted from \citep{b17}, wherein the enhanced electron density model by \citet{b1} was used to estimate the average altitude. Thus 17 GHz seems to originate almost at the interface of the chromosphere and the corona, hence, the equatorial rotation rate matches more with the photospheric or the chromospheric data rather than the coronal data. The analysis of daily solar full disc radio images at 17 GHz for the years 1999 to 2001 shows that the coronal equatorial rotation rate is higher than those obtained by XBPs and quite lower than those obtained by CBPs \citep{b11}.

The coefficient $B$ shows that the corona does rotate differentially as in the photosphere and chromosphere; but the gradient is lower ($<$ 25.8\% and $<$31.3\%, respectively). This is in agreement with the results obtained by \citet{b18} and \citet{b19}. The difference in the rotation profiles of soft X-ray emission and 17 GHz could possibly be due to their different regions of origin in the solar atmosphere. The coefficients $A$ and $B$ are also compared with the annual sunspot numbers (see Table ~1). This gives evidence of the dependence of coronal rotation on the phases of solar cycle.

Figure~5 shows that the coronal equatorial rotation ($A$) is in phase with the annual sunspot number, whereas its latitude dependent differential gradient is anti-phase to the annual sunspot number. A long term study is indeed required to ascertain this correlation more clearly.

\section*{Acknowledgments}

The authors wish to acknowledge the data (radio images at 17 GHz \& annual sunspot numbers) used in the present work. These are acquired from the web page of {\it Nobeyama Radio Observatory\/} (NRO) \& {\it National Geophysical Data Centre} (NGDC), respectively. We are indebted to the observers who were involved in the acquisition of the useful solar data. The research at {\it Physical Research Laboratory\/} (PRL) is supported by {\it Department of Space\/}, Government of India. The authors (SC and KNI) are also thankful to their parent institutes for their support. The authors thank the anonymous referee for their valuable suggestions on the manuscript.


\bsp

\label{lastpage}

\end{document}